\documentclass[showpacs,preprintnumbers,showkeys,amsmath,amssymb]{revtex4}
\usepackage{graphicx}
\usepackage{dcolumn}
\usepackage{bm}
\begin{document}

\preprint{}

\title{Implementing two-photon three-degree-of-freedom hyper-parallel controlled phase flip gate through cavity-assisted interactions}

\author{Hai-Rui Wei,\footnote{Corresponding author:hrwei@ustb.edu.cn}  Wen-Qiang Liu, and Ning-Yang Chen}

\address{School of Mathematics and Physics, University of Science and Technology Beijing, Beijing 100083, China}

\begin{abstract}
Hyper-parallel quantum information processing is a promising and beneficial research field. In this paper, we present a method to implement a hyper-parallel controlled-phase-flip (hyper-CPF) gate for frequency-, spatial-, and time-bin-encoded qubits by coupling flying photons to trapped nitrogen vacancy (NV) defect centers. The scheme, which differs from their conventional parallel counterparts, is specifically advantageous in decreasing against the dissipate noise, increasing the quantum channel capacity, and reducing the quantum resource overhead. The gate qubits with frequency, spatial, and time-bin degrees of freedom (DOF) are immune to quantum decoherence in optical fibers, whereas the polarization photons are easily disturbed by the ambient noise.
\end{abstract}



\maketitle

\section{Introduction}\label{sec1}

In recent years, great advancements have been made in many areas of quantum information processing (QIP), including quantum teleportation \cite{teleportation1,teleportation2}, quantum secret sharing \cite{sharing}, quantum key distribution \cite{QKD2,QKD3}, quantum secure direct communication \cite{QSDC1,QSDC2}, quantum dense coding \cite{dense}, quantum algorithms \cite{algorithm1,algorithm2,algorithm3,algorithm4}, and quantum gates \cite{gate1,gate2,gate4}. Because quantum communication utilizes quantum coherent superposition and quantum entanglement effect, its propagation rate and reliability are higher than those of conventional communication methods \cite{book}. Further, quantum computing exhibits a higher performance than its conventional counterparts to efficiently search the target items in an unsorted date base and factor large integers \cite{book}. Recently, numerous sophisticated approaches have been proposed to improve the conventional methods by employing multiple degrees of freedom (DOFs). Multiple DOFs are beneficial for a wide range of applications, including the implementation of hyper-parallel quantum computation \cite{hyper}, quantum communication \cite{analysis2}, simplification of quantum computation \cite{simplify}, high-dimension QIP \cite{high-dimension2}, and completion of specific deterministic tasks that cannot be solved by single DOF systems, such as deterministic linear optical quantum algorithms \cite{DJ2}, deterministic linear optical quantum gates \cite{sigle-photon-three}, linear optical teleportation \cite{teleportation2}, and quantum key  distribution   without  a  shared  reference  frame \cite{deterministic-QKD}. Furthermore, hyper-parallel QIP  has been gaining great attention because of its excellent advantages, making it can interesting and potential candidate for long distance quantum secure communication and quantum computers.


Hyper-parallel QIP, whose operations are simultaneously executed in two or more distinct DOFs, is potentially  robust against photonic dissipation noise, and enhances the quantum channel capacity, improves the security of quantum communication, reduces experimental requirements and resource overheads, augments the success rate of protocols, and improves the speed of quantum computation.
Recently, various hyper-entangled states have been reported; for example,  polarization-spatial-energy hyper-entangled states \cite{polarization-spatial-energy}, polarization-time-bin hyper-entangled states \cite{polarization-time}, spin-motion hyper-entangled states  \cite{spin-motion},  polarization-momentum hyper-entangled states \cite{polarization-momentum}, polarization-time-frequency hyper-entangled states \cite{polarization-time-frequency}, and multiple-path hyper-entangled states \cite{multiple-path}. These resources can help us implement many important quantum tasks with one DOF, such as completing entangled states analysis using linear optics \cite{BSA1,BSA3}, entanglement purification and concentration \cite{purification}, one-DOF cluster state preparation and one way quantum computing \cite{one-way}, quantum error-correcting \cite{error-correcting}, teleportation \cite{polarization-momentum}, linear photonic superdense coding \cite{super-dense-coding2}, enhanced violation of local realism \cite{violation-local-realism1}, and quantum algorithm \cite{multiple-path}. Moreover, hyper-entanglement has provided other important applications in hyper-parallel photonic quantum computing \cite{hyper-gate5,hyper-gate6}, hyper-entangled swapping \cite{hyper-swapping}, hyper-teleportation \cite{hyperteleportation}, hyper-entangled states analysis \cite{HBSA2,HBSA3,HBSA4}, hyper-parallel repeater \cite{hyper-repeater}, hyper-entanglement purification \cite{hyper-purification1, hyper-purification3}, and hyper-entanglement concentration \cite{hyper-concentration5,hyper-concentration6}.


Photons have emerged as excellent candidates for hyper-parallel QIP because of their vast amount of available qubits, such as DOFs, including polarization \cite{gate-polarization}, spatial mode \cite{polarization-spatial-energy}, transverse orbital angular momentum \cite{orbital1,orbital2}, time-bin \cite{time-bin1}, frequency (or color) \cite{gate-frequency1}, and continuous-variable energy-time modes  \cite{energy-time}. Furthermore, not only can photons easily carry quantum information over long distances due to negligible decoherence in free space, but they can also be manipulated in an extremely fast and precise way by linear optical elements and produced in a highly efficient way \cite{SPDC}. Flexible control of photons by employing standard linear optics has been an interesting approach to implement probabilistic photonic QIP \cite{gate1}. It has been demonstrated that the deterministic QIP can be achieved by encoding computing qubits in different DOFs in a photonic architecture. However, scalability is a major challenge in this approach because it only allows for  one photon system. Currently, cross-Kerr medium, atoms, atom assemble, and artificial atoms (such as quantum dot, superconductor, and diamond nitrogen vacancy defect center) are often employed as a prominent platform to address the scalability problem that overcomes the intrinsic weak interactions between individual photons. In recent years, diamond nitrogen vacancy (NV) in defect centers has gained great attention because of its exceptional features, such as excellent scalability, optical property, and ultra-long coherence time even at room temperature \cite{temperature1,temperature2}. Electronic spin confined in NV center \cite{Ai1,Ai2} enables generation of spin-photon entanglement; in addition, assistance using  NV center, polarization-spatial hyper-parallel quantum computing, hyperentanglement purification, hyperentanglement concentration, and hyperentangled state analysis have been proposed.  Cavity quantum electrodynamics significantly enhances photon-matter interactions, and it is essential for quantum networks and distributed quantum computation \cite{Hu2008,PRX}. Conditional transmitters/reflecters of the NV-cavity platform have a wide range of applications both in the strong-coupling regime of the high-$Q$ cavity and the weak-coupling region of the low-$Q$ cavity. Nowadays, conditional transmission and reflection techniques based on photon-NV interactions have been extensively employed for QIP ranging from implementations of quantum gates \cite{gate01,gate02} to measurement-device-independent quantum key distribution \cite{QKD01,QKD02}, quantum networks \cite{network1,network2}, hyper-parallel quantum gates \cite{hyper-gate01,hyper-gate02}, and multiple DOFs entanglement distribution \cite{multi-entanglement-distribution}.



In this paper, we design a scheme to efficiently implement optical hyper-parallel controlled phase flip (hyper-CPF) gate; a hyper-parity gate simultaneously acts on frequency, spatial, and time-bin DOFs through cavity-assisted interactions. The CPF gate or similar logic operations are essential for quantum communication and quantum computation, and our gate mechanism is deterministic in principle. It is note that polarization photons are highly susceptible to decoherence in optical fibers  because of  polarization  mode  dispersion caused by thermal fluctuation, vibration, and imperfections of the fiber \cite{polarization-mode-dispersion1,polarization-mode-dispersion2}. On the contrary, frequency, spatial, and time-bin photons are immune to these decoherence effects. Furthermore, frequency photons can efficiently transfer quantum information at telecommunication wavelengths  \cite{frequency-advantage2,frequency-advantage3,frequency-advantage4}, and time-bin photons can minimize the effects of detector dead-time and have the advantage of relative insensitivity to inhomogeneities in transmission media \cite{time-bin-robust1,time-bin-robust2,time-bin-robust3}. Moreover, in contrast to conventional quantum computation with one DOF, our constructions are not only more robust against the photonic dissipation noise, but can also increase the capacity of the quantum channel and reduce the demand for quantum resources.


\begin{figure} 
\centering
\includegraphics[width=9.5 cm,angle=0]{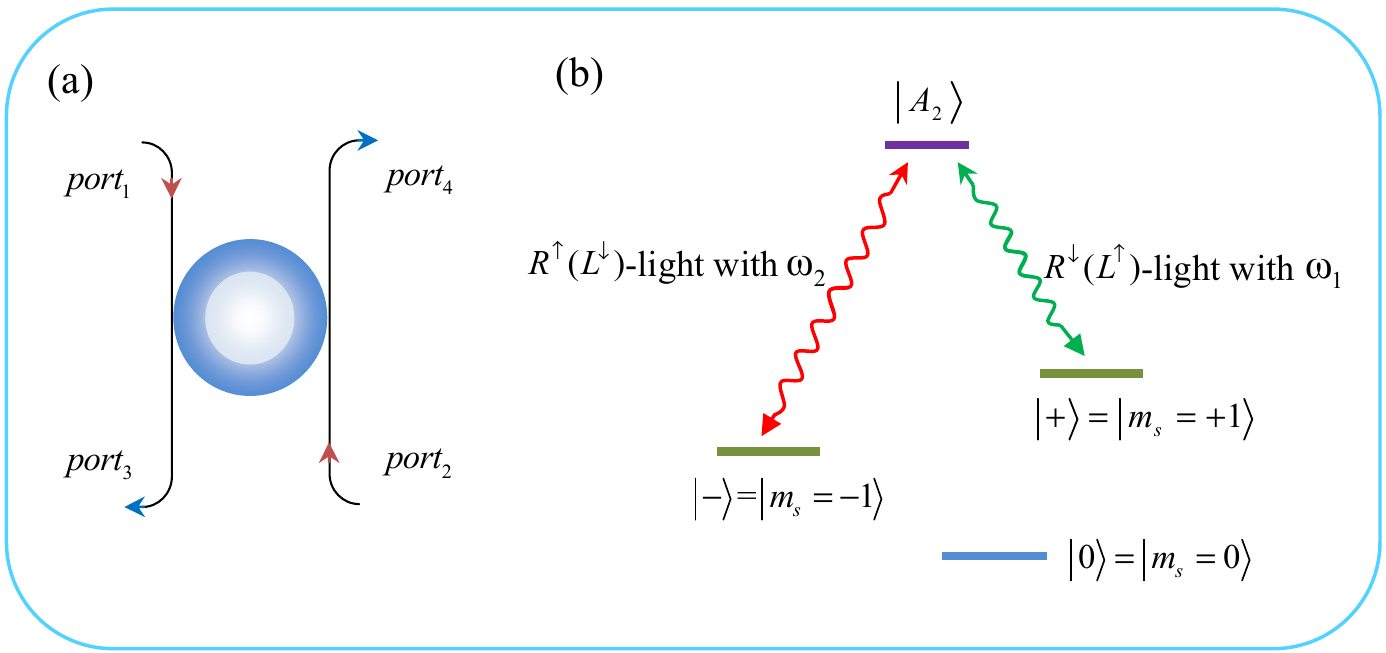}
\caption{(Color online) (a) Schematic of a diamond NV defect center trapped in the double-sided resonant microcavity. (b) Schematic of an NV center coupling to a resonator and the possible $\Lambda$-type optical transitions in an NV center. } \label{level}
\end{figure}


\section{Hyper-CPF gate and hyper-parity-check}\label{sec2}

\textbf{The optical property of a double-sided NV-cavity platform.}\quad
In recent years, constant efforts have been made toward the diamond NV center. As shown in Fig. \ref{level}, the electron-spin triple ground states of a negatively charged NV$^-$ center are split into $|m_{s}=0\rangle$ (marked as $|0\rangle$), and $|m_{s}=\pm1\rangle$ (marked as $|\pm\rangle$) by 2.88 GHz with zero  magnetic field  due to the crystal field \cite{spin}.  The degeneracy levels $|\pm\rangle$  can be further split according to Zeeman effect by applying a magnetic field ($2\pi \times 200$ MHz) to the sample along one of the NV axes \cite{Parameter}.
The six excited states \cite{excited}
$|A_{1}\rangle=(|E_{-}\rangle|+\rangle-|E_{+}\rangle|-\rangle)/\sqrt{2}, |A_{2}\rangle=(|E_{-}\rangle|+\rangle+|E_{+}\rangle)|-\rangle)/\sqrt{2},
|E_{1}\rangle=(|E_{-}\rangle|-\rangle-|E_{+}\rangle|+\rangle)/\sqrt{2}, |E_{2}\rangle=(|E_{-}\rangle|-\rangle+|E_{+}\rangle|+\rangle)/\sqrt{2},
|E_{x}\rangle=|X\rangle|0\rangle$, and $|E_{y}\rangle=|Y\rangle|0\rangle$
are dominated by the $\rm C_{3v}$ symmetry of the NV center, spin-spin, spin-orbit interactions without external strain, and electric or magnetic fields. $|E_{\pm}\rangle, |X\rangle$, and $|Y\rangle$ represent the orbital states of a diamond NV center with the angular momentum projections $\pm1$ and 0, respectively.
It is known that spin-orbit interaction splits the pair ($A_1$, $A_2$) away from the others by about 5.5 GHz, and  the spin-spin interaction separates states $|A_{1}\rangle$ and $|A_{2}\rangle$ by about 3.3 GHz \cite{split1,split2}. Thus, for the small strain and magnetic fields, the optically excited state $|A_{2}\rangle$ is robust with stable symmetric properties and preserves the polarization properties of its optical transitions to states $|\pm\rangle$  through polarized $|\sigma_\mp\rangle$  radiations due to total angular momentum conservation \cite{excited}. The state $|A_{1}\rangle$  is coupled to the  meta-stable singlet state  $|^1A_1\rangle$ and then decays to the ground state $|0\rangle$ \cite{meta-stable}. Hence, states  $|A_{2}\rangle$  and $|\pm\rangle$  provide the most ideal the  $\Lambda$-type three-level system for QIP. That is, as shown in Fig. \ref{level}(b),
the right- (left-) circularly-polarized photons (labeled as $|R^{\downarrow}\rangle$ and $|L^{\uparrow}\rangle$) with the frequency $\omega_{1}$ only strongly couple $|+\rangle$  to $|A_{2}\rangle$, whereas the right- (left-) circularly-polarized photons (labeled as $|R^{\uparrow}\rangle$ and $|L^{\downarrow}\rangle$) with frequency $\omega_{2}$ only strongly couple $|-\rangle$ to $|A_{2}\rangle$; here, the superscript $\uparrow$ ($\downarrow$) of polarized photons $R$ and $L$ represents the propagation directions of the photons, which are along (against) the $z$ axis (i.e., NV axis).

The reflection/transmission coefficients of the cavities can be obtained by solving Heisenberg equations of motion for the cavity field operator $\hat{a}$ and the negatively charged exciton $ X^-$ dipole operator $\sigma_-$ \cite{Heisenberg,Heisenberg2},
\begin{eqnarray}          \label{eq2}
&&\frac{d\hat{a}}{dt}=-\left[i(\omega_c-\omega)+\kappa+\frac{\kappa_s}{2}\right]\hat{a}-\text{g}\sigma_{-}-\sqrt{\kappa}\,\hat{a}_{in}-\sqrt{\kappa}\,\hat{a}_{in}'+\hat{H},\nonumber\\
&&\frac{d\sigma_-}{dt}=-\left[i(\omega_{X^-}-\omega)+\frac{\gamma}{2}\right]\sigma_{-}-\text{g}\sigma_z\hat{a}+\hat{G},
\end{eqnarray}
and the input--output relations of the cavity \cite{Heisenberg}
\begin{equation}        \label{eq3}
\hat{a}_r=\hat{a}_{in}+ \sqrt{\kappa}\,\hat{a},\qquad\qquad
\hat{a}_t=\hat{a}_{in}'+\sqrt{\kappa}\,\hat{a}.
\end{equation}
Here, $\omega$, $\omega_c$, and $\omega_{X^-}$ are the frequencies of the incident single photon, cavity mode, and $ X^-$ dipole transition, respectively; g denotes the coupling constant of the $X^-$-cavity combination; $\gamma/2$, $\kappa$, and $\kappa_s/2$ are the decay rates of the $ X^-$ dipole, cavity field, and side leakage, respectively;  $\hat{H}$ and $\hat{G}$ are the noise operators; $\sigma_z$ represents the Pauli operator; $\hat{a}_{in}$ ($ \hat{a}'_{in}$) and $\hat{a}_t$  ($\hat{a}_{r}$) denote the input and output field operators, respectively.
When $ X^-$ stays predominantly in the ground state, i.e., taking $\langle\sigma_z\rangle\approx-1$, the reflection/transmission coefficients $r(\omega)$/$t(\omega)$ of the diamond NV center can be calculated as \cite{transmission1},
\begin{eqnarray}             \label{eq4}
&& r(\omega)=\frac{[i(\omega_{X^{-}}-\omega)+\frac{\gamma}{2}][i(\omega_c-\omega)+\frac{\kappa_s}{2}]+g^2}{\left[i(\omega_{X^{-}}-\omega)+\frac{\gamma}{2}\right]\left[i(\omega_c-\omega)+\kappa+\frac{\kappa_s}{2}\right]+g^2},\nonumber\\
&& t(\omega)=\frac{-\kappa\left[i(\omega_{X^{-}}-\omega)+\frac{\gamma}{2}\right]}{\left[i(\omega_{X^{-}}-\omega)+\frac{\gamma}{2}\right]\left[i(\omega_c-\omega)+\kappa+\frac{\kappa_s}{2}\right]+g^2}.
\end{eqnarray}
By adjusting the single-photon pulse to resonate with the cavity mode and the $ X^-$ dipole transition ($\omega=\omega_c=\omega_{X^-}$), and considering the ideal conditions that the Purcell factor $ g^{2}/(\kappa\gamma)\gg1$ with  $\kappa_{s}\approx0$ in the hot cavity mode ($g\neq0$), one can obtain $r(\omega)\rightarrow1$ and $t(\omega)\rightarrow0$;  $\kappa_{s}/\kappa\ll1$ with  $\kappa_{s}\approx0$ in the cold cavity mode ($g=0$), one can obtain\emph{} $ r_{0}(\omega)\rightarrow0$ and $t_{0}(\omega)\rightarrow-1$. That implicit mechanism is that when the photon strongly couples to the hot cavity mode, it is reflected by the hot cavity without any phase shift; when the coupling strength $g$ and $\kappa_{s}$ are negligible, the transmitted photon experiences a $\pi$--phase shift. Thus, the spin-dependence optical transition rules can be summarized as,
\begin{eqnarray}             \label{eq5}
&&|R^{\uparrow},\omega_{1},+\rangle\rightarrow -|R^{\uparrow},\omega_{1},+\rangle, \quad\qquad |R^{\downarrow},\omega_{1},+\rangle\rightarrow |L^{\uparrow},\omega_{1},+\rangle, \nonumber\\
&&|R^{\uparrow},\omega_{2},+\rangle\rightarrow -|R^{\uparrow},\omega_{2},+\rangle, \quad\qquad |R^{\downarrow},\omega_{2},+\rangle\rightarrow -|R^{\downarrow},\omega_{2},+\rangle, \nonumber\\
&&|R^{\uparrow},\omega_{1},-\rangle\rightarrow -|R^{\uparrow},\omega_{1},-\rangle, \quad\qquad |R^{\downarrow},\omega_{1},-\rangle\rightarrow -|R^{\downarrow},\omega_{1},-\rangle, \nonumber\\
&&|R^{\uparrow},\omega_{2},-\rangle\rightarrow |L^{\downarrow},\omega_{2},-\rangle, \quad\;\;\,\,\qquad |R^{\downarrow},\omega_{2},-\rangle\rightarrow -|R^{\downarrow},\omega_{2},-\rangle, \nonumber\\
&&|L^{\uparrow},\omega_{1},+\rangle\rightarrow |R^{\downarrow},\omega_{1},+\rangle, \quad\;\;\,\,\qquad|L^{\downarrow},\omega_{1},+\rangle\rightarrow -|L^{\downarrow},\omega_{1},+\rangle,\nonumber\\
&&|L^{\uparrow},\omega_{2},+\rangle\rightarrow -|L^{\uparrow},\omega_{2},+\rangle, \quad\qquad\,|L^{\downarrow},\omega_{2},+\rangle\rightarrow -|L^{\downarrow},\omega_{2},+\rangle,\nonumber\\
&&|L^{\uparrow},\omega_{1},-\rangle\rightarrow -|L^{\uparrow},\omega_{1},-\rangle, \quad\qquad\,|L^{\downarrow},\omega_{1},-\rangle\rightarrow -|L^{\downarrow},\omega_{1},-\rangle, \nonumber\\
&&|L^{\uparrow},\omega_{2},-\rangle\rightarrow -|L^{\uparrow},\omega_{2},-\rangle, \quad\qquad\,|L^{\downarrow},\omega_{2},-\rangle\rightarrow |R^{\uparrow},\omega_{2},-\rangle.
\end{eqnarray}

The total spontaneous emission rate $\gamma_{\text{total}}=2\pi \times15$ MHz, of the NV center was demonstrated in 2010 \cite{excited}. In 2009, Barclay \emph{et al.} \cite{microdisks} experimentally demonstrated the NV center relevant
parameters [$g_{\text{ZPL}}$, $\kappa$, $\gamma_{\text{total}}$, $\gamma_{\text{ZPL}}]/2\pi =[0.30, 26, 0.013, 0.0004]$ GHz. Here, $g_{\text{ZPL}}$ is the coupling strength  of the single microdisk photon and the NV zero phonon line (ZPL), $\gamma_{\text{total}}$ and $\gamma_{\text{ZPL}}$ are the total and ZPL spontaneous optical decay rates of the NV center, respectively. Fortunately, $\gamma_{\text{ZPL}}/\gamma_{\text{total}}$ has been enhanced from the 3-4\%
to 70\% \cite{Enhanced-ZPL}.  In the following, we will employ the coherence emission within the narrow-band ZPL described by Eq. (\ref{eq5}) to construct hyper-CPF gate.

\begin{figure} 
\centering
\includegraphics[width=12.5 cm,angle=0]{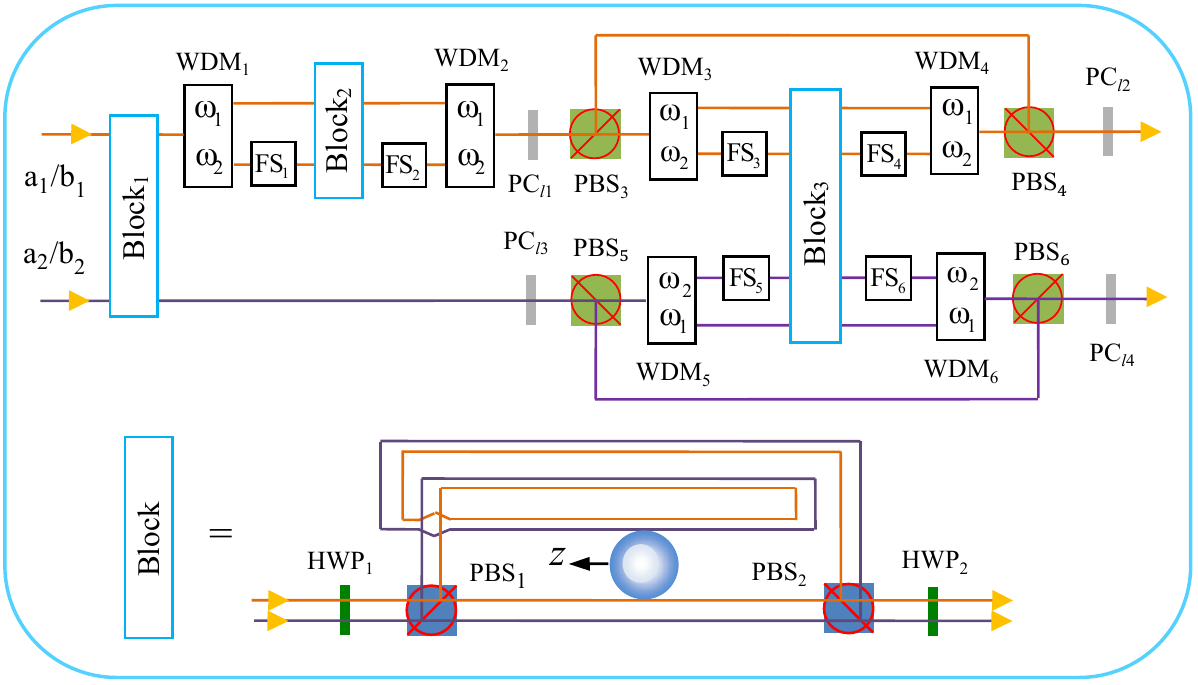}
\caption{(Color online) A schematic for implementing the two-photon three-DOF hyper-CPF gate.
PBS$_{i}$ ($i=1,\;2,\; \cdots,\; 6$) are circularly polarizing beam splitters that transmit the $R$-polarized photon and reflect the $L$-polarized photon.
WDM$_{i}$ ($i=1,\; 2,\; \cdots,\; 6$) represent the polarization independent wavelength division multiplexers, which lead the photons to different spatial modes according to their frequencies.
FS$_{i}$ ($i=1,\; 2,\; \cdots,\; 6$) are frequency shifters, which are used to complete the qubit flip operation on the frequency of a single photon, i.e., $\omega_{1}\leftrightarrow\omega_{2}$.
The Pockels cells $\text{PC}_{li}$ ($i=1,\; 2,\; 3,\; 4$) perform bit-flip operations on the polarization DOF of the photons when the $l$-time-bin component appears.
The half-wave plates $\rm HWP_{1}$ and $\rm HWP_{2}$ are oriented at $\rm22.5^{\circ}$ to complete the Hadamard transformations.} \label{CPF}
\end{figure}

\textbf{Implementation of the hyper-CPF gate}.
Based on the above photon-NV platform, we design a quantum circuit to a two-photon three-DOF CPF gate in which the gate qubits are independently encoded in the frequency, spatial, and time-bin DOFs of the single photon system (see Fig. \ref{CPF}). The electron spins confined in the NV centers act as ancilla qubits.

Suppose that the initial states of the photons $a$ and $b$, and the three electronic spins $e_1$, $e_2$, and $e_3$ in diamond NV centers are
\begin{eqnarray}             \label{eq6}
&&|\varphi\rangle_a=|R\rangle_{a}(\alpha_{1}|\omega_{1}\rangle_{a}+\alpha_{2}|\omega_{2}\rangle_{a})(\gamma_{1}|a_{1}\rangle_{a}+\gamma_{2}|a_{2}\rangle_{a})
(\varsigma_{1}|l_{1}\rangle_{a}+\varsigma_{2}|s_{2}\rangle_{a}),\nonumber \\
&&|\varphi\rangle_b=|R\rangle_{b}(\beta_{1}|\omega_{1}\rangle_{b}+\beta_{2}|\omega_{2}\rangle_{b})(\delta_{1}|b_{1}\rangle_{b}+\delta_{2}|b_{2}\rangle_{b})
(\xi_{1}|l_{1}\rangle_{b}+\xi_{2}|s_{2}\rangle_{b}),\nonumber\\
&&|\varphi\rangle_{e_1}=\frac{1}{\sqrt{2}}(|+\rangle_{1}+|-\rangle_{1}), \quad
|\varphi\rangle_{e_2}=\frac{1}{\sqrt{2}}(|+\rangle_{2}+|-\rangle_{2}), \quad
|\varphi\rangle_{e_3}=\frac{1}{\sqrt{2}}(|+\rangle_{3}+|-\rangle_{3}).
\end{eqnarray}
Here,
$|\omega_1\rangle_a$ and $|\omega_2\rangle_a$ ($|\omega_1\rangle_b$ and $|\omega_2\rangle_b$),
$|a_1\rangle_a$ and $|a_2\rangle_a$ ($|b_1\rangle_b$ and $|b_2\rangle_b$),
$|l_1\rangle_a$ and $|s_2\rangle_a$ ($|l_1\rangle_b$ and $|s_2\rangle_b$)
are the two frequency-qubit states, spatial-qubit states, and  time-bin-qubit states of photon $a$ ($b$), respectively; $|R\rangle_a$ ($|R\rangle_b$) denotes the $R$-polarized state of the photon $a$ ($b$). The complex coefficients $\alpha_{i}$, $\gamma_{i}$, $\varsigma_{i}$, $\beta_{i}$, $\delta_{i}$, and $\xi_{i}$  satisfy the normalization condition $|\alpha_{1}|^{2}+|\alpha_{2}|^{2}=1$, $|\gamma_{1}|^{2}+|\gamma_{2}|^{2}=1$, $|\varsigma_{1}|^{2}+|\varsigma_{2}|^{2}=1$, $|\beta_{1}|^{2}+|\beta_{2}|^{2}=1$, $|\delta_{1}|^{2}+|\delta_{2}|^{2}=1$, and $|\xi_{1}|^{2}+|\xi_{2}|^{2}=1$, respectively.

Firstly, the photon $a$ is injected; then, it interacts with ``Block$_1$'' composed of HWP$_1$, PBS$_1$, NV, PBS$_2$, and HWP$_2$. The polarizing beam splitters PBS$_1$ and PBS$_2$ transmit the $R$-polarized photon and reflect the $L$-polarized photon, respectively. The half-wave plates HWP$_1$ and HWP$_2$ are rotated to 22.5$^\circ$ to complete the Hadamard transformations as follows:
\begin{eqnarray}             \label{eq7}
|R\rangle\xrightarrow{H_p}\frac{1}{\sqrt{2}}(|R\rangle+|L\rangle), \quad
|L\rangle\xrightarrow{H_p}\frac{1}{\sqrt{2}}(|R\rangle-|L\rangle).
\end{eqnarray}
Combining the above-mentioned facts and  Eq. (\ref{eq5}), it can be seen that the ``Block'' completes the following transformations:
\begin{eqnarray}             \label{eq8}
&&|R^{\downarrow},\omega_{1}\rangle|+\rangle\xrightarrow{\rm Block}|R^{\downarrow},\omega_{1}\rangle|+\rangle,\qquad
 |R^{\downarrow},\omega_{1}\rangle|-\rangle\xrightarrow{\rm Block}-|R^{\downarrow},\omega_{1}\rangle|-\rangle,\nonumber\\
&&|R^{\downarrow},\omega_{2}\rangle|+\rangle\xrightarrow{\rm Block}-|R^{\downarrow},\omega_{2}\rangle|+\rangle,\quad\;
 |R^{\downarrow},\omega_{2}\rangle|-\rangle\xrightarrow{\rm Block}-|R^{\downarrow},\omega_{2}\rangle|-\rangle,\nonumber\\
&&|L^{\uparrow},\omega_{1}\rangle|+\rangle\xrightarrow{\rm Block}-|L^{\uparrow},\omega_{1}\rangle|+\rangle,\quad\;\;
 |L^{\uparrow},\omega_{1}\rangle|-\rangle\xrightarrow{\rm Block}-|L^{\uparrow},\omega_{1}\rangle|-\rangle,\nonumber\\
&&|L^{\uparrow},\omega_{2}\rangle|+\rangle\xrightarrow{\rm Block}-|L^{\uparrow},\omega_{2}\rangle|+\rangle,\quad\;\;
 |L^{\uparrow},\omega_{2}\rangle|-\rangle\xrightarrow{\rm Block}-|L^{\uparrow},\omega_{2}\rangle|-\rangle.
\end{eqnarray}
Therefore, the ``Block$_1$'' ($\rm HWP_{1}\rightarrow PBS_1\rightarrow$NV$\rm\rightarrow PBS_2\rightarrow HWP_{2})$ evolves the state of the entire system $\rm|\varphi\rangle_0$ into $\rm|\varphi\rangle_1$. Here,
\begin{eqnarray}                  \label{eq9}
|\varphi\rangle_0=|\varphi\rangle_a \otimes |\varphi\rangle_b \otimes |\varphi\rangle_{e_1} \otimes |\varphi\rangle_{e_2} \otimes |\varphi\rangle_{e_3},
\end{eqnarray}
\begin{eqnarray}                  \label{eq10}
&&|\varphi\rangle_1=\frac{1}{\sqrt{2}}|R^{\downarrow}\rangle_{a}[\alpha_{1}|\omega_{1}\rangle_{a}(|+\rangle_{1}-|-\rangle_{1})-\alpha_{2}|\omega_{2}\rangle_{a}(|+\rangle_{1}+|-\rangle_{1})]
\nonumber\\&&\qquad\quad\otimes (\gamma_{1}|a_{1}\rangle_{a}+\gamma_{2}|a_{2}\rangle_{a})\otimes(\varsigma_{1}|l_{1}\rangle_{a}+\varsigma_{2}|s_{2}\rangle_{a})
\otimes|\varphi\rangle_b \otimes |\varphi\rangle_{e_2} \otimes |\varphi\rangle_{e_3}.
\end{eqnarray}
After Hadamard operations $H_es$ are performed on the NV center $e_1$, the state $|\varphi\rangle_1$ becomes
\begin{eqnarray}                  \label{eq11}
&&|\varphi\rangle_2=|R^{\downarrow}\rangle_{a}[\alpha_{1}|\omega_{1}\rangle_{a}|-\rangle_{1}-\alpha_{2}|\omega_{2}\rangle_{a}|+\rangle_{1}]
\otimes (\gamma_{1}|a_{1}\rangle_{a}+\gamma_{2}|a_{2}\rangle_{a})\otimes(\varsigma_{1}|l_{1}\rangle_{a}+\varsigma_{2}|s_{2}\rangle_{a})
\otimes|\varphi\rangle_b \otimes |\varphi\rangle_{e_2} \otimes |\varphi\rangle_{e_3}.
\end{eqnarray}
Here, $H_es$ executes the following operations:

\begin{eqnarray}             \label{eq12}
|+\rangle\xrightarrow{H_e}\frac{1}{\sqrt{2}}(|+\rangle+|-\rangle), \quad
|-\rangle\xrightarrow{H_e}\frac{1}{\sqrt{2}}(|+\rangle-|-\rangle).
\end{eqnarray}

Secondly, a polarization independent wavelength division multiplexer WDM$_1$ separates the wavepackets emitted from spatial model $a_1$ into two arms $a_{1\rm U}$ and $a_{1\rm D}$ according to their frequencies. And a frequency shifter FS$_1$ flips the frequency of the photon emitted from the spatial $a_{\rm 1D}$, i.e., $|\omega_1\rangle|a_{\rm 1D}\rangle\leftrightarrow|\omega_2\rangle|a_{\rm 1D}\rangle$. Subsequently, the wavepackets converge at WDM$_2$. The pockels cells PC$_{l1}$ and PC$_{l3}$ perform bit-flip operations on the polarization DOF of the passing photon when the $l$-time-bin component occurs. The PBS$_3$ (PBS$_5$) directly reflects the $L$-polarization components to PBS$_4$ (PBS$_6$) and transmits the $R$-polarization components to the operations consisting of  WDM$_3$, FS$_3$,  ``Block$_3$'',  FS$_4$,   WDM$_4$ and PBS$_4$ (WDM$_5$, FS$_5$,  ``Block$_3$'',  FS$_6$,  WDM$_6$ and PBS$_6$).  Thus, these operations WDM$_1$, FS$_1$, ``Block$_2$'', FS$_2$,   WDM$_2$, PC$_{l1}$,  PBS$_3$,  WDM$_3$,  FS$_3$,  ``Block$_3$'',  FS$_4$, WDM$_4$, PBS$_4$, $H_{e_3}$ and PC$_{l2}$ (PC$_{l3}$, PBS$_5$,  WDM$_5$,  FS$_5$,  ``Block$_3$'',  FS$_6$, WDM$_6$, PBS$_6$, $H_{e_3}$ and PC$_{l4}$) change the state $|\varphi\rangle_2$ into $|\varphi\rangle_3$,
\begin{eqnarray}                  \label{eq14}
&&|\varphi\rangle_{3}=|R^{\downarrow}\rangle_{a}(\alpha_{1}|-\rangle_{1}|\omega_{1}\rangle_{a}-\alpha_{2}|+\rangle_{1}|\omega_{2}\rangle_{a})(\gamma_{1}|-\rangle_2|a_{1}\rangle_{a} + \gamma_{2}|+\rangle_2|a_{2}\rangle_{a})(\varsigma_{1}|l_{1}\rangle_{a}|+\rangle_3 + \varsigma_{2}|s_{2}\rangle_{a}|-\rangle_3) \otimes|\varphi\rangle_b.
\end{eqnarray}

Thirdly, the photon $b$ is injected, and similar operations as those for the photon $a$ are performed. That is, the photon $b$ passes through the ``Block$_1$'', $\rm WDM_{1}$, $\rm FS_{1}$, ``Block$_2$'', FS$_2$,  WDM$_2$, PC$_{l1}$, PBS$_3$, WDM$_3$, FS$_3$,  ``Block$_3$'',  FS$_4$, WDM$_4$, PBS$_4$ and  PC$_{l2}$ (``Block$_1$'', $\rm WDM_{1}$, $\rm FS_{1}$, ``Block$_2$'', FS$_2$,  WDM$_2$, PC$_{l3}$, PBS$_5$, WDM$_5$, FS$_5$,  ``Block$_3$'',  FS$_6$, WDM$_6$, PBS$_6$ and  PC$_{l4}$), and the $H_e$s are applied on three NV centers, which cause the state of the system becomes
\begin{eqnarray}          \label{eq16}
&&|\varphi\rangle_{4}=|R\rangle_{a}|R\rangle_{b}\{[-\alpha_{1}|\omega_{1}\rangle_{a}(\beta_{1}|\omega_{1}\rangle_{b}+\beta_{2}|\omega_{2}\rangle_{b})+\alpha_{2}|\omega_{2}\rangle_{a}(\beta_{2}|\omega_{2}\rangle_{b}-\beta_{1}|\omega_{1}\rangle_{b})]|+\rangle_{1}\nonumber\\
&&\qquad\quad+[\alpha_{1}|\omega_{1}\rangle_{a}(\beta_{1}|\omega_{1}\rangle_{b}+\beta_{2}|\omega_{2}\rangle_{b})+\alpha_{2}|\omega_{2}\rangle_{a}(\beta_{2}|\omega_{2}\rangle_{b}-\beta_{1}|\omega_{1}\rangle_{b})]|-\rangle_{1}\}\nonumber\\
&&\qquad\quad\otimes\{[\gamma_{1}|a_{1}\rangle_{a}(\delta_{2}|b_{2}\rangle_{b}-\delta_{1}|b_{1}\rangle_{b})+\gamma_{2}|a_{2}\rangle_{a}(\delta_{1}|b_{1}\rangle_{b}+\delta_{2}|b_{2}\rangle_{b})]|+\rangle_{2}\nonumber\\
&&\qquad\quad+[\gamma_{1}|a_{1}\rangle_{a}(\delta_{1}|b_{1}\rangle_{b}-\delta_{2}|b_{2}\rangle_{b})+\gamma_{2}|a_{2}\rangle_{a}(\delta_{1}|b_{1}\rangle_{b}+\delta_{2}|b_{2}\rangle_{b})]|-\rangle_{2}\}\nonumber\\
&&\qquad\quad\otimes\{[\varsigma_{1}|l_{1}\rangle_{a}(\xi_{1}|l_{1}\rangle_{b}+\xi_{2}|s_{2}\rangle_{b})+\varsigma_{2}|s_{2}\rangle_{a}(\xi_{1}|l_{1}\rangle_{b}-\xi_{2}|s_{2}\rangle_{b})]|+\rangle_{3}\nonumber\\
&&\qquad\quad+[\varsigma_{1}|l_{1}\rangle_{a}(\xi_{1}|l_{1}\rangle_{b}+\xi_{2}|s_{2}\rangle_{b})+\varsigma_{2}|s_{2}\rangle_{a}(\xi_{2}|s_{2}\rangle_{b}-\xi_{1}|l_{1}\rangle_{b})]|-\rangle_{3}\}.
\end{eqnarray}

\begin{table}[!htb]
\centering
\caption{The corresponding relations between the feed-forward operations applied  on the frequency, spatial and time-bin DOFs, and the spin states of the $\rm NV_{1,2,3}$ centers. $I_2$ is an identical transformation and $\sigma_z$ is a $\pi$-phase shift on $|\omega_{2}\rangle$, $|b_{2}\rangle$ or $|s_{2}\rangle$ and $|\omega_{1}\rangle$, $|b_{1}\rangle$, $|s_{1}\rangle$ remain unchanged.}
\begin{tabular}{lccccc}

\hline  \hline

          & \multicolumn {1}{c}{Feed--forward} \\
\hline
\;\;    &   control photon ($a$)      &   target photon ($b$)     \\

$|+\rangle_{1}$ (frequency)   &  $-I_2$         &  $I_2$           \\

$|-\rangle_{1}$ (frequency)   &  $\sigma_z$         & $I_2$        \\

$|+\rangle_{2}$ (spatial)   &  $-\sigma_z$         & $\sigma_z$      \\

$|-\rangle_{2}$ (spatial)   &  $I_2$           & $\sigma_z$      \\

$|+\rangle_{3}$ (time-bin)   &  $I_2$            & $I_2$         \\

$|-\rangle_{3}$ (time-bin)   &  $\sigma_z$         & $I_2$         \\

                             \hline  \hline
\end{tabular} \label{Tab1}
\end{table}

Finally, we measure the outcomes of the three NV spins based on \{$\rm|\pm\rangle$\} basis, and perform some feed-forward operations on the exiting photons according to Tab. \ref{Tab1}. A deterministic two-photon hyper-CPF gate with frequency, spatial, and time-bin DOFs is completed, in which phase flips are applied to the elements $|\omega_2\rangle_a|\omega_2\rangle_b$ (frequency), $|a_2\rangle_a|b_2\rangle_b$ (spatial), and $|s_2\rangle_a|s_2\rangle_b$ (time-bin), whereas the others remain unchanged. This hyper-CPF gate is equivalent to the hyper-parallel controlled-NOT (hyper-CNOT) gate up to two local Hadamard transformations.

\textbf{Deterministic hyper-parity gate}.
The parity gate is the key to implementing entanglement swapping and quantum repeater, measuring the inequality of the Bell, designing quantum algorithms, constructing quantum gates, and entanglement purification and concentration \cite{parity-check1,parity-check2,parity-check3,parity-check5,parity-check6}.

Fig. \ref{CPF} can additionally perform hyper-parity gate in the frequency, spatial-mode, and time-bin DOFs simultaneously. Further, as described in Fig. \ref{CPF}, after the photon $a$ is injected, the initial state $|\varphi\rangle_{0}$ of the entire system  is transformed into the state $|\varphi\rangle_{3}$. Before and after the photon $b$ passes through the ``Block$_1$'', ``Block$_2$'' and ``Block$_3$'' the $H_e$s are executed on three electron-spins successively, which transforms the state $|\varphi\rangle_{3}$ into
\begin{eqnarray}          \label{eq19}
&&|\varphi\rangle'_{4}=|R\rangle_{a}|R\rangle_{b}[(\alpha_{1}\beta_{1}|\omega_{1}\rangle_{a}|\omega_{1}\rangle_{b}+\alpha_{2}\beta_{2}|\omega_{2}\rangle_{a}|\omega_{2}\rangle_{b})|+\rangle_{1}-(\alpha_{1}\beta_{2}|\omega_{1}\rangle_{a}|\omega_{2}\rangle_{b}\nonumber\\
&&\qquad\quad+\alpha_{2}\beta_{1}|\omega_{2}\rangle_{a}|\omega_{1}\rangle_{b})|-\rangle_{1}]\otimes[(\delta_{1}\gamma_{1}|a_{1}\rangle_{a}|b_{1}\rangle_{b}+\delta_{2}\gamma_{2}|a_{2}\rangle_{a}|b_{2}\rangle_{b})|+\rangle_{2}\nonumber\\
&&\qquad\quad+(\delta_{1}\gamma_{2}|a_{2}\rangle_{a}|b_{1}\rangle_{b}+\delta_{2}\gamma_{1}|a_{1}\rangle_{a}|b_{2}\rangle_{b})|-\rangle_{2}]\otimes[(\varsigma_{1}\xi_{1}|l_{1}\rangle_{a}|l_{1}\rangle_{b}\nonumber\\
&&\qquad\quad+\varsigma_{2}\xi_{2}|s_{2}\rangle_{a}|s_{2}\rangle_{b})|+\rangle_{3}+(\varsigma_{2}\xi_{1}|s_{2}\rangle_{a}|l_{1}\rangle_{b}+\varsigma_{1}\xi_{2}|l_{1}\rangle_{a}|s_{2}\rangle_{b})|-\rangle_{3}].
\end{eqnarray}

Therefore, on detecting the electronic spin states $|+\rangle_{1,2,3}$, the joint state $|\varphi\rangle'_{4}$ collapses into the even hyper-parity-state
\begin{eqnarray}          \label{eq20}
(\alpha_{1}\beta_{1}|\omega_{1}\rangle_{a}|\omega_{1}\rangle_{b}+\alpha_{2}\beta_{2}|\omega_{2}\rangle_{a}|\omega_{2}\rangle_{b})\otimes(\delta_{1}\gamma_{1}|a_{1}\rangle_{a}|b_{1}\rangle_{b}+\delta_{2}\gamma_{2}|a_{2}\rangle_{a}|b_{2}\rangle_{b})\otimes(\varsigma_{1}\xi_{1}|l_{1}\rangle_{a}|l_{1}\rangle_{b}+\varsigma_{2}\xi_{2}|s_{2}\rangle_{a}|s_{2}\rangle_{b}). \end{eqnarray}
On detecting the electronic spin states $|-\rangle_{1,2,3}$, the joint state $|\varphi\rangle'_{4}$ collapses into the odd hyper-parity-state
\begin{eqnarray}          \label{eq21}
(\alpha_{1}\beta_{2}|\omega_{1}\rangle_{a}|\omega_{2}\rangle_{b}+\alpha_{2}\beta_{1}|\omega_{2}\rangle_{a}|\omega_{1}\rangle_{b})\otimes(\delta_{1}\gamma_{2}|a_{2}\rangle_{a}|b_{1}\rangle_{b}+\delta_{2}\gamma_{1}|a_{1}\rangle_{a}|b_{2}\rangle_{b})\otimes(\varsigma_{2}\xi_{1}|s_{2}\rangle_{a}|l_{1}\rangle_{b}+\varsigma_{1}\xi_{2}|l_{1}\rangle_{a}|s_{2}\rangle_{b}).
\end{eqnarray}
For $|+\rangle_{1,2}$ and $|-\rangle_{3}$, it collapses into
\begin{eqnarray}          \label{eq22}
(\alpha_{1}\beta_{1}|\omega_{1}\rangle_{a}|\omega_{1}\rangle_{b}+\alpha_{2}\beta_{2}|\omega_{2}\rangle_{a}|\omega_{2}\rangle_{b})\otimes(\delta_{1}\gamma_{1}|a_{1}\rangle_{a}|b_{1}\rangle_{b}+\delta_{2}\gamma_{2}|a_{2}\rangle_{a}|b_{2}\rangle_{b})\otimes(\varsigma_{2}\xi_{1}|s_{2}\rangle_{a}|l_{1}\rangle_{b}+\varsigma_{1}\xi_{2}|l_{1}\rangle_{a}|s_{2}\rangle_{b}). \end{eqnarray}
For $|+\rangle_{1,3}$ and $|-\rangle_{2}$, it collapses into
\begin{eqnarray}          \label{eq23}
(\alpha_{1}\beta_{1}|\omega_{1}\rangle_{a}|\omega_{1}\rangle_{b}+\alpha_{2}\beta_{2}|\omega_{2}\rangle_{a}|\omega_{2}\rangle_{b})\otimes(\delta_{1}\gamma_{2}|a_{2}\rangle_{a}|b_{1}\rangle_{b}+\delta_{2}\gamma_{1}|a_{1}\rangle_{a}|b_{2}\rangle_{b})\otimes(\varsigma_{1}\xi_{1}|l_{1}\rangle_{a}|l_{1}\rangle_{b}+\varsigma_{2}\xi_{2}|s_{2}\rangle_{a}|s_{2}\rangle_{b}). \end{eqnarray}
For $|+\rangle_{1}$ and $|-\rangle_{2,3}$, it collapses into
\begin{eqnarray}          \label{eq24}
(\alpha_{1}\beta_{1}|\omega_{1}\rangle_{a}|\omega_{1}\rangle_{b}+\alpha_{2}\beta_{2}|\omega_{2}\rangle_{a}|\omega_{2}\rangle_{b})\otimes(\delta_{1}\gamma_{2}|a_{2}\rangle_{a}|b_{1}\rangle_{b}+\delta_{2}\gamma_{1}|a_{1}\rangle_{a}|b_{2}\rangle_{b})\otimes(\varsigma_{2}\xi_{1}|s_{2}\rangle_{a}|l_{1}\rangle_{b}+\varsigma_{1}\xi_{2}|l_{1}\rangle_{a}|s_{2}\rangle_{b}). \end{eqnarray}
For $|-\rangle_{1}$ and $|+\rangle_{2,3}$, it collapses into
\begin{eqnarray}          \label{eq25}
(\alpha_{1}\beta_{2}|\omega_{1}\rangle_{a}|\omega_{2}\rangle_{b}+\alpha_{2}\beta_{1}|\omega_{2}\rangle_{a}|\omega_{1}\rangle_{b})\otimes(\delta_{1}\gamma_{1}|a_{1}\rangle_{a}|b_{1}\rangle_{b}+\delta_{2}\gamma_{2}|a_{2}\rangle_{a}|b_{2}\rangle_{b})\otimes(\varsigma_{1}\xi_{1}|l_{1}\rangle_{a}|l_{1}\rangle_{b}+\varsigma_{2}\xi_{2}|s_{2}\rangle_{a}|s_{2}\rangle_{b}). \end{eqnarray}
For $|-\rangle_{1,3}$ and $|+\rangle_{2}$, it collapses into
\begin{eqnarray}          \label{eq26}
(\alpha_{1}\beta_{2}|\omega_{1}\rangle_{a}|\omega_{2}\rangle_{b}+\alpha_{2}\beta_{1}|\omega_{2}\rangle_{a}|\omega_{1}\rangle_{b})\otimes(\delta_{1}\gamma_{1}|a_{1}\rangle_{a}|b_{1}\rangle_{b}+\delta_{2}\gamma_{2}|a_{2}\rangle_{a}|b_{2}\rangle_{b})\otimes(\varsigma_{2}\xi_{1}|s_{2}\rangle_{a}|l_{1}\rangle_{b}+\varsigma_{1}\xi_{2}|l_{1}\rangle_{a}|s_{2}\rangle_{b}).
\end{eqnarray}
For $|-\rangle_{1,2}$ and $|+\rangle_{3}$, it collapses into
\begin{eqnarray}          \label{eq27}
(\alpha_{1}\beta_{2}|\omega_{1}\rangle_{a}|\omega_{2}\rangle_{b}+\alpha_{2}\beta_{1}|\omega_{2}\rangle_{a}|\omega_{1}\rangle_{b})\otimes(\delta_{1}\gamma_{2}|a_{2}\rangle_{a}|b_{1}\rangle_{b}+\delta_{2}\gamma_{1}|a_{1}\rangle_{a}|b_{2}\rangle_{b})\otimes(\varsigma_{1}\xi_{1}|l_{1}\rangle_{a}|l_{1}\rangle_{b}+\varsigma_{2}\xi_{2}|s_{2}\rangle_{a}|s_{2}\rangle_{b}).
\end{eqnarray}
It is noted that Eqs. (\ref{eq22}-\ref{eq27}) can be transformed into the Eq. (\ref{eq20}) and Eq. (\ref{eq21}) by performing single-qubit feed-forward operations.


\section{Discussion and Summary}\label{sec3}

Optical QIP has been extensively studied in recent years, and previous studies are mainly focused on single DOF of the photon. The KLM scheme \cite{gate1} was a cornerstone in linear optical quantum computing with significant success probability. Matter qubits, ranging from cross-Kerr to natural and artificial atoms (quantum dot, superconducting, NV center in diamond), are often employed to ensure deterministic gate reciprocity between isolated individual photons. Great progress has been made in such entangled photon-matter platform.

However, the giant Kerr nonlinear remains a challenge in experiments. Further, millions neutral atom can trapped in microscopic arrays, long coherence time of neutral atoms can be achieved at very low temperatures ($n$K-$\mu$K), while individual manipulation and readout of neutral atoms in optical lattices are not possible \cite{atom}. Superconducting qubits have $\mu$s-scale coherence time and operate at $m$K temperature \cite{superconductor}. A semiconductor quantum dot circuit operates at a few K and supports $\mu$s-scale coherence time \cite{QD-coherence}. The NV center in diamond has an ultra-long coherence time ($\sim$ms) \cite{coherence-time2}; moreover, it could operate even at room temperature \cite{temperature1,temperature2}. The initialization \cite{initializtion}, manipulation ($\sim$ subnanosecond) \cite{coherence1,subnanosecond2}, and readout ($\rm\sim$100 $\mu$s) \cite{readout1,readout2} of the electronic spin in the NV center have been experimentally demonstrated.  The interactions between the NV centers and the various optical microcavities including microspheres \cite{microsphere}, microdisks \cite{microdisks}, microrings \cite{microrings}, microtoroidal resonators \cite{resonators}, photonic crystals \cite{crystals} and fiber-based microcavities \cite{fiber-cavity} have been demonstrated in the  realistic experiments.
To date, photon-NV \cite{hybrid} and NV-NV \cite{NV-NV1,NV-NV2} entangled states and their applications have been reported.

Our schemes allow for the polarization-degenerate microcavities. Fortunately, polarization-degenerate microcavities have been experimentally demonstrated in H1 photonic crystal cavities \cite{H1-photonic-crystal,H1-photonic-crystal2}, micropillar cavities \cite{micropillar,micropillar2}, and  fiber-based cavities \cite{fiber-cavity}.
In our work, the NV centers trapped in double-sided microcavities act as ancilla qubit for implementing hyper-CPF gate and hyper-parity gate. Unlike the conventional single DOF CPF gates, our gates simultaneously operate on frequency and spatial and time-bin DOFs of the photon.  Tunable single-frequency-qubit operations \cite{frequency-operation} and arbitrary linear-optics time-bin encoding single-qubit operations \cite{time-operation} have been demonstrated; arbitrary spatial single-qubit operations can be completed by using beam splitters, half-wave plates and quarter-wave plates \cite{BS} and spatial modes can also been extracted from a light beam without disturbance to other orthogonal modes \cite{spatial}.
Compared with the polarization schemes, our scheme is robust against decoherence in optical fibers \cite{polarization-mode-dispersion1,polarization-mode-dispersion2,frequency-advantage4,time-bin-robust1,time-bin-robust2,frequency-advantage2,frequency-advantage3}.
Moreover, our three-DOF hyper-CPF gate is superior to its one-DOF and two-DOF counterparts in reducing the resource overhead, improving the success rate and the quantum channel capacity, increasing the operation speed, and reducing the experimental requirements; in addition, it is highly robust against the environment. Additional photons, which are necessary for cross-Kerr-based \cite{cross-kerr} and parity-check-based gates \cite{parity-check1}, are not required in our scheme. The double-sided-cavity-based construction is more robust and flexible than the single-sided-cavity-based ones because it does not require the balance of coefficients for the coupled and the uncoupled cavity to achieve high fidelity \cite{hu2009}.


Our schemes are constructed under the ideal conditions $ t_{0}(\omega)\approx-1$, $r_{0}(\omega)\approx0$, $ r(\omega) \approx 1$,  $ t(\omega)\approx0$, i.e., side leakage and physical imperfections in the construction processes are not considered. However, the inevitable nonzero photon bandwidth, mismatch, and the finite coupling rate between the photon and the cavity mode induce imperfect birefringence of the cavity, and the side leakage of the cavity reduce the performance of our scheme. Thus, the interactions between the incident photon pulse and the NV center in Eq. (\ref{eq5}) should be rewritten as,
\begin{eqnarray}             \label{eq28}
&&|R^{\uparrow},\omega_{1},+\rangle\rightarrow t_{0}|R^{\uparrow},\omega_{1},+\rangle+r_{0}|L^{\downarrow},\omega_{1},+\rangle, \;\;\, |R^{\downarrow},\omega_{1},+\rangle\rightarrow t|R^{\downarrow},\omega_{1},+\rangle+r|L^{\uparrow},\omega_{1},+\rangle, \nonumber\\
&&|R^{\uparrow},\omega_{2},+\rangle\rightarrow t_{0}|R^{\uparrow},\omega_{2},+\rangle+r_{0}|L^{\downarrow},\omega_{2},+\rangle, \;\;\, |R^{\downarrow},\omega_{2},+\rangle\rightarrow t_{0}|R^{\downarrow},\omega_{2},+\rangle+r_{0}|L^{\uparrow},\omega_{2},+\rangle, \nonumber\\
&&|R^{\uparrow},\omega_{1},-\rangle\rightarrow t_{0}|R^{\uparrow},\omega_{1},-\rangle+r_{0}|L^{\downarrow},\omega_{1},-\rangle, \;\;\, |R^{\downarrow},\omega_{1},-\rangle\rightarrow t_{0}|R^{\downarrow},\omega_{1},-\rangle+r_{0}|L^{\uparrow},\omega_{1},-\rangle, \nonumber\\
&&|R^{\uparrow},\omega_{2},-\rangle\rightarrow t|R^{\uparrow},\omega_{2},-\rangle+r|L^{\downarrow},\omega_{2},-\rangle, \;\;\;\;\;\,
|R^{\downarrow},\omega_{2},-\rangle\rightarrow t_{0}|R^{\downarrow},\omega_{2},-\rangle+r_{0}|L^{\uparrow},\omega_{2},-\rangle, \nonumber\\
&&|L^{\uparrow},\omega_{1},+\rangle\rightarrow t|L^{\uparrow},\omega_{1},+\rangle+r|R^{\downarrow},\omega_{1},+\rangle, \;\;\;\;\;\;
|L^{\downarrow},\omega_{1},+\rangle\rightarrow t_{0}|L^{\downarrow},\omega_{1},+\rangle+r_{0}|R^{\uparrow},\omega_{1},+\rangle,\nonumber\\
&&|L^{\uparrow},\omega_{2},+\rangle\rightarrow t_{0}|L^{\uparrow},\omega_{2},+\rangle+r_{0}|R^{\downarrow},\omega_{2},+\rangle,\;\;\;
|L^{\downarrow},\omega_{2},+\rangle\rightarrow t_{0}|L^{\downarrow},\omega_{2},+\rangle+r_{0}|R^{\uparrow},\omega_{2},+\rangle,\nonumber\\
&&|L^{\uparrow},\omega_{1},-\rangle\rightarrow t_{0}|L^{\uparrow},\omega_{1},-\rangle+r_{0}|R^{\downarrow},\omega_{1},-\rangle,\;\;\;
|L^{\downarrow},\omega_{1},-\rangle\rightarrow t_{0}|L^{\downarrow},\omega_{1},-\rangle+r_{0}|R^{\uparrow},\omega_{1},-\rangle,\nonumber\\
&&|L^{\uparrow},\omega_{2},-\rangle\rightarrow t_{0}|L^{\uparrow},\omega_{2},-\rangle+r_{0}|R^{\downarrow},\omega_{2},-\rangle, \;\;\;
|L^{\downarrow},\omega_{2},-\rangle\rightarrow t|L^{\downarrow},\omega_{2},-\rangle+r|R^{\uparrow},\omega_{2},-\rangle.
\end{eqnarray}
To evaluate the performance of the hyper-CPF gate and the hyper-parity gate, we simulated the results of the average fidelity and efficiency of the ``Block''. It is known that the fidelity of arbitrary two quantum states $\rho$ and $\sigma$ is defined as \cite{book}
$F(\rho,\sigma)\equiv tr\sqrt{\rho^{1/2}\sigma\rho^{1/2}}=\sqrt{\langle \rho|\sigma|\rho\rangle}
$. Let $|\varphi\rangle$ and $|\phi\rangle$ ($\sigma=|\phi\rangle\langle\phi|$) be purification chosen such that $F(\rho,\sigma)=|\langle\varphi|\phi\rangle|=F(|\varphi\rangle,|\phi\rangle)$.
Therefore, the  fidelity of ``Block'' operation can be defined as  $F=|\langle\psi_{\text{real}}|\psi_{\text{ideal}}\rangle|$,  in which $|\psi_{\text{real}}\rangle$ and $|\psi_{\text{ideal}}\rangle$ represent the terminal states of the system after the ``Block'' operation in realistic and ideal conditions, respectively. The ``Block'' efficiency is defined as $\eta=n_{\text{output}}/n_{\text{input}}$. Here, $n_{\text{input}}$ and $n_{\text{output}}$ are the numbers of input and output photons in the construction of the ``Block'', respectively.

\begin{figure}  [ht]        
\centering
\includegraphics[width=7.5 cm, angle=0]{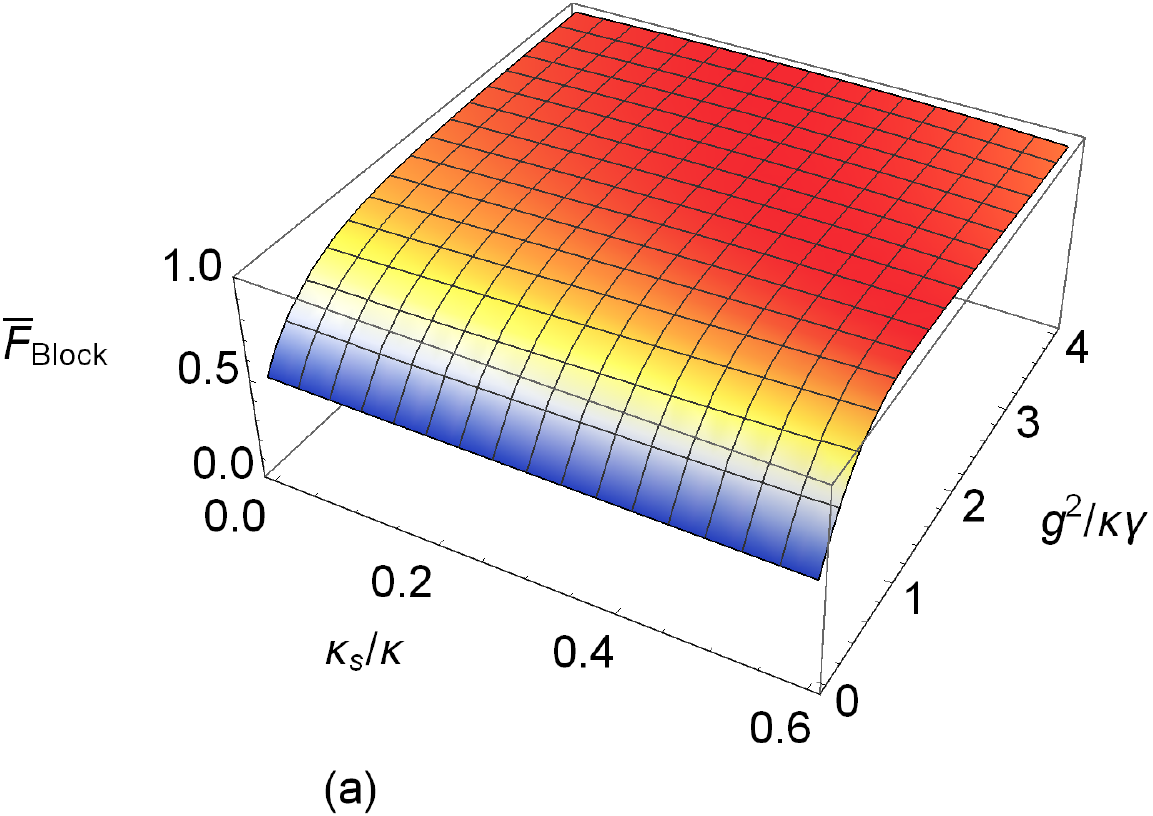}
\includegraphics[width=7.5 cm, angle=0]{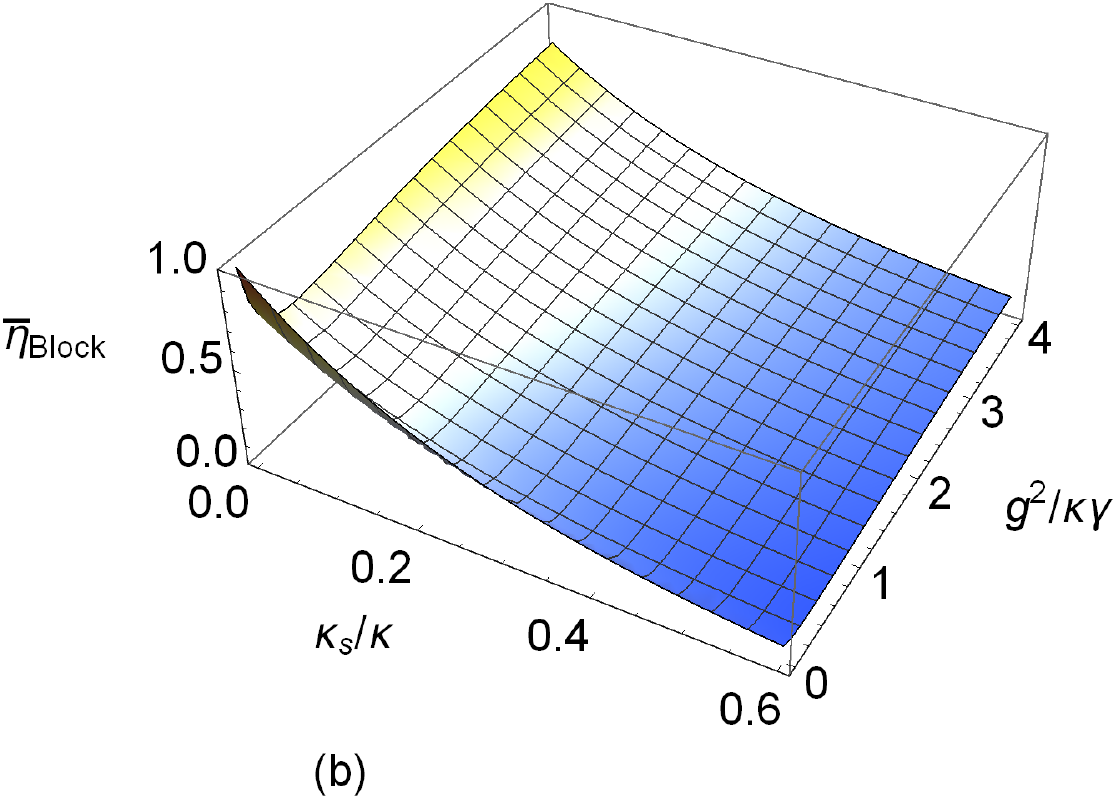}\\
\caption{(Color online)   The average fidelity and efficiency of the ``Block'' operation vs  leakage rate $\kappa_s/\kappa$ and the parameter  $g^2/\kappa\gamma$. (a) The average fidelity of the ``Block''; (b) The average efficiency of the ``Block''. The resonator condition $\omega_c=\omega_{X^-}=\omega$ is taken.} \label{Fidelity-Efficiency}
\end{figure}

Combining the gate construction processes and Eq. (\ref{eq28}), one can find that the arbitrary initial state ($|\psi_{\text{init}}\rangle_{\text{Block}}$) and the terminal states ($|\psi_{\text{ideal}}\rangle_{\text{Block}}$ and  $|\psi_{\text{real}}\rangle_{\text{Block}}$) of the ``Block'' applied on two photons can be written as
\begin{eqnarray}
&&|\psi_{\text{init}}\rangle_{\text{Block}}=\frac{1}{\sqrt{2}}|R\rangle_a|R\rangle_b
(\cos\alpha|\omega_1\rangle_a+\sin\alpha|\omega_2\rangle_a)
(\cos\beta|\omega_1\rangle_b+\sin\beta|\omega_2\rangle_b)
(|+\rangle+|-\rangle),\\
%
&&|\psi_{\text{ideal}}\rangle_{\text{Block}}=\frac{1}{\sqrt{2}}|R\rangle_a|R\rangle_b
(\cos\alpha\cos\beta |\omega_1\rangle_a|\omega_1\rangle_b|+\rangle
-\cos\alpha\sin\beta |\omega_1\rangle_a|\omega_2\rangle_b|+\rangle\nonumber\\&&\qquad\qquad\qquad\;
+\cos\alpha\cos\beta |\omega_1\rangle_a|\omega_1\rangle_b|-\rangle
+\cos\alpha\sin\beta |\omega_1\rangle_a|\omega_2\rangle_b|-\rangle\nonumber\\&&\qquad\qquad\qquad\;
-\sin\alpha\cos\beta |\omega_2\rangle_a|\omega_1\rangle_b|+\rangle
+\sin\alpha\sin\beta |\omega_2\rangle_a|\omega_2\rangle_b|+\rangle\nonumber\\&&\qquad\qquad\qquad\;
+\sin\alpha\cos\beta |\omega_2\rangle_a|\omega_1\rangle_b|-\rangle
+\sin\alpha\sin\beta |\omega_2\rangle_a|\omega_2\rangle_b|-\rangle),\\
%
&&|\psi_{\text{real}}\rangle_{\text{Block}}=\frac{1}{\sqrt{2}}|R\rangle_a|R\rangle_b
(\cos\alpha\cos\beta (t+r)^2       |\omega_1\rangle_a|\omega_1\rangle_b|+\rangle
+\cos\alpha\cos\beta (t_0+r_0)^2   |\omega_1\rangle_a|\omega_1\rangle_b|-\rangle\nonumber\\&&\qquad\qquad\qquad\;
+\sin\alpha\cos\beta (t_0+r_0)(t+r)|\omega_2\rangle_a|\omega_1\rangle_b|+\rangle
+\sin\alpha\cos\beta (t_0+r_0)^2   |\omega_2\rangle_a|\omega_1\rangle_b|-\rangle\nonumber\\&&\qquad\qquad\qquad\;
+\cos\alpha\sin\beta (t_0+r_0)(t+r)|\omega_1\rangle_a|\omega_2\rangle_b|+\rangle
+\cos\alpha\sin\beta (t_0+r_0)^2   |\omega_1\rangle_a|\omega_2\rangle_b|-\rangle\nonumber\\&&\qquad\qquad\qquad\;
+\sin\alpha\sin\beta (t_0+r_0)^2   |\omega_2\rangle_a|\omega_2\rangle_b|+\rangle
+\sin\alpha\sin\beta(t_0+r_0)^2    |\omega_2\rangle_a|\omega_2\rangle_b|-\rangle).
\end{eqnarray}
By averaging over $\alpha,\; \beta\in[0,\;2\pi]$, the average fidelity and efficiency $\overline{F}_{\text{Block}}$ and $\overline{\eta}_{\text{Block}}$ can be expressed as
\begin{eqnarray}          \label{eq29}
&&\overline{F}_{\text{Block}}=\frac{1}{(2\pi)^2}\int_{0}^{2\pi}d\alpha\int_{0}^{2\pi}d\beta|\langle\psi_{\text{real}}|\psi_{\text{ideal}}\rangle|,\\ &&\overline{\eta}_{\text{Block}}=\frac{1}{(2\pi)^2}\int_{0}^{2\pi}d\alpha\int_{0}^{2\pi}d\beta\frac{n_{\text{output}}}{n_{\text{input}}}=\frac{(r+t)^4+2(r+t)^2(r_0+t_0)^2+5(r_0+t_0)^4}{8}.
\end{eqnarray}
According to the realistic NV parameters \cite{microdisks}, one can obtain $g^2/\kappa\gamma=8.654$ in which $r(\omega)\approx0.95$ for the resonant condition $\omega=\omega_c=\omega_{X^-}$; for the larger parameter $g^2/\kappa\gamma\geq25$ with $Q\sim10^5$ (corresponding to $\kappa\sim1$ GHz) or $Q\sim10^4$ (corresponding to $\kappa\sim10$ GHz), $r(\omega)$ can reach nearly unity \cite{Parameter}. The cavity side leakage rate $10\kappa_s\approx\kappa$ can be achieved with the current state of the art NV-cavity fabrication techniques \cite{ks/k,ks/k1}.
The average fidelity and efficiency of the ``Block'' operation as functions of $\kappa_s/\kappa$ and $g^2/\kappa\gamma$ are depicted in Fig. \ref{Fidelity-Efficiency}. One can find that the NV-cavity coupling strength, cavity side leakage, and imperfect birefringence have a great impact on average fidelity and efficiency. High average fidelity and efficiency can be achieved by increasing $g^2/\kappa\gamma$ and decreasing $\kappa_s/\kappa$.  When $\kappa_s/\kappa=0.1$ and $g^2/\kappa\gamma=8.654$, the average fidelity is $\overline{F}_{\text{Block}}$=99.99\% and the efficiency is $\overline{\eta}_{\text{Block}}$=66.01\%.

In actuality, the major experimental imperfections that reduce the fidelity of the photon-NV block are: (a) imperfections in electronic spin manipulations and readout \cite{readout-fidelity,control-fidelity}, such as the electronic spin preparation (reduction  $\rm<1\%$), spin decoherence (reduction  $\rm<1\%$), offresonant excitation errors (reduction  $\rm\approx1\%$), spin-flip errors in the excited states (reduction  $\rm\approx1\%$), microwave pulse errors (reduction  $\rm\approx3.5\%$) \cite{readout1,hybrid,fidelity3}, and  detector dark counts (reduction 3\%); (b) spatial mode mismatch between cavity and incident photon (reduction 3\%); (c) stability of the differences between the cavity resonance and the frequency of the incident photon; (d) small probability of two photons in one qubit mode pulses (reduction 2\%); (e) errors induced by optical elements, such as PBS, HWP, and optical fiber; (f) linear optical elements and fiber absorption losses. However, these limitations are not fundamental.


To summarize, we designed a compact quantum circuit to implement an two-photon three-DOF hyper-CPF gate by utilizing the significant advantages, such as frequency, spatial, and time-bin DOFs of photons. The hyper-CPF gate mechanism based on the photon-NV entangled building block is deterministic and robust against photonic dissipation noise because the gate is hyperparallel and the computing qubits are encoded in multiple DOFs. Moreover, our hyper-CPF gate not only increased the quantum channel capacity but also reduced the quantum resources overhead. Moreover, as an interesting application, we proposed the hyper-parity gate for hyper-parallel quantum computing and hyper-parallel quantum communication. Finally, the evaluations indicated that our schemes are feasible with the current experimental technology.

\bigskip

\section*{Acknowledgements}

The work is supported by the National Natural Science Foundation of China under Grant No. 11604012, and the Fundamental Research Funds for the Central Universities under Grant Nos. FRF-TP-19-011A3, 230201506500024 and FRF-BR-17-004B, and a grant from China Scholarship Council.


\end{document}